# The Dawn of Mathematical Biology

*Daniel Sander Hoffmann*
*Computer Engineering Unit. UERGS, RS, Brazil.*
*E-mail: daniel-hoffmann@uergs.edu.br*

**Abstract**
In this paper I describe the early development of the so-called mathematical biophysics, as conceived by Nicolas Rashevsky back in the 1920´s, as well as his latter idealization of a "relational biology". I also underline that the creation of the journal "The Bulletin of Mathematical Biophysics" was instrumental in legitimating the efforts of Rashevsky and his students, and I finally argue that his pioneering efforts, while still largely unacknowledged, were vital for the development of important scientific contributions, most notably the McCulloch-Pitts model of neural networks.

## 1. Introduction

The modern era of theoretical biology can be classified into "foundations," "physics and chemistry," "cybernetics" and "mathematical biophysics" (Morowitz, 1965). According to this author, an important part of the history of the modern era in theoretical biology dates back to the beginning of the twentieth century, with the publication of D'Arcy Wentworth Thompson's opus "On Growth and Form" (Thompson, 1917), closely followed by works like the "Elements of Physical Biology" of Alfred J. Lotka (1925). Some authors tracked Lotka's ideas closely, yielding books such as "Leçons sur la Théorie Mathématique de la Lutte pour la Vie," by Vito Volterra (1931), and Kostitzin's "Biologie Mathématique" (Kostitzin, 1937). Mathematics and ecology do share a long coexistence, and mathematical ecology is currently one of the most developed areas of the theoretical sciences taken as a whole. Genetics is another example of great success in modern applied mathematics, its history beginning (at least) as early as in the second decade of the last century, when J. B. S. Haldane published "A Mathematical Theory of Natural and Artificial Selection" (Haldane, 1924), followed by "The Genetical Theory of Natural Selection," in 1930, and "The Theory of Inbreeding," in 1949, both by R. A. Fisher (see Fisher, 1930, 1949). The approach of "physics and chemistry" is represented by workers like Erwin Schrödinger – one of the founding fathers of quantum mechanics –, who wrote a small but widely read book entitled "What is Life?" (Schrödinger, 1944), and Hinshelwood (1946), with his "The Chemical Kinetics of the Bacterial Cell." "Cybernetics" is a successful term coined by Norbert Wiener and used in a book with the same name (Wiener, 1948). At the same time C. E. Shannon (1948) published his seminal paper "A Mathematical Theory of Communication," and it is a known fact that both works profoundly influenced a whole generation of mathematical biologists and other theoreticians. While Wiener stressed the importance of feedback – with the notion of closed loop control yielding new approaches to theoretical biology, ecology and the neurosciences –, information theory was improved and applied in several technological and scientific areas. Finally, the amalgamation of information theory with the notion of feedback strongly influenced the work of important theoretical ecologists like Robert E. Ulanowicz (1980, 1997) and Howard T. Odum (1983).

I suggest that the development of the last division highlighted above, namely "mathematical biophysics" is, up to now, largely unknown to mainstream historians and philosophers of science. Interestingly enough, this unfamiliarity spreads even to most historians and philosophers of biology. However, I wish to point out a recent revival of some fundamental ideas associated with this school, a fact that, alone, justifies a closer look into the origins of this investigative framework. Accordingly, in this paper I review and briefly discuss some early stages of this line of thought.

## 2. The roots of mathematical biophysics and of the relational approach

Nicolas Rashevsky was born in Chernigov on September 1899 (Cull, 2007). He took a Ph.D. in theoretical physics very early in his life, and soon began publishing in quantum theory and relativity, among other topics. He immigrated to North America in 1927, after being trained in Russia as a mathematical physicist. His original work in biology began when he moved to the Research Laboratories of the Westinghouse Corporation in Pittsburgh, Pennsylvania, where he worked on the thermodynamics of liquid droplets. There he found that these structures became unstable past a given critical size, spontaneously dividing into smaller droplets. Later still, while involved with the Mathematical Biology program of the University of Chicago, Rashevsky studied cell division and excitability phenomena. The Chicago group established "The Bulletin of Mathematical Biophysics" (now "The Bulletin of Mathematical Biology"), an important contribution to the field of theoretical biology. In this journal one finds most of Rashevsky's published biological material, and it also served to introduce the work of many of his students. Thus, the journal helped to catapult new careers and (above all) catalyze the formation and maintenance of the "mathematical biophysics" school, still an influential school in modern theoretical biology (Rosen, 1991). To be fair, the journal was widely open to all interested researchers. More than that, Rashevsky and his colleague Herb Landahl used to take for themselves the task of correcting and even helping to extend the mathematics, encouraging the authors to re-submit the papers. As a side note, it is interesting to mention that, given the shortcomings of the publication of graphics at the time, Landahl offered invaluable help to the authors, carefully preparing each drawing for printing (Cull, 2007).

The importance of organizing new journals, proceedings and books for "legitimating" a new branch of science was emphasized by Smocovitis (1996). I would like to suggest, therefore, that the "mathematical biophysics" case fits nicely this interpretation. Other periodicals of importance that arose during this period include "Acta Biotheoretica," founded in 1935 by the group then at the Professor Jan der Hoeven Foundation for Theoretical Biology of the University of Leiden, "Bibliographia Biotheoretica" (published by the same group) and the well-known "Journal of Theoretical Biology," founded in 1961 (Morowitz, 1965).

Rashevsky is rightly acknowledged for the proposition of a systematic approach to the use of mathematical methods in biology. He intended to develop a "mathematical biology" that would relate to experiments just like the well-established mathematical physics (Cull, 2007). He chose to name this new field of inquiry "mathematical biophysics," a decision reflected in the title of the aforementioned journal. By the mid-1930s Rashevsky had already explored the links between chemical reactions and physical diffusion (currently known as "reaction-diffusion" phenomena), as well as the associated destabilization of homogeneous states that is at the core of the modern notion of self-organization (Rosen, 1991). An elaborated theory of cell division based on the principles behind diffusion drag forces was offered close to the end of that decade (Rashevsky, 1939), and led to new equations concerning the rates of constriction and elongation of demembranated *Arbacia* eggs under division (Landahl, 1942a, 1942b). (I note that *Arbacia* is a genus of hemispherically-shaped sea urchins commonly used in experiments of the kind.) This theory agreed well with previously available empirical data, but Rashevsky himself urged to demonstrate that the theory of diffusion drag forces was inadequate to represent most other facts of cell division known at the time (Rosen, 1991).

Rashevsky's major work is (rather unsurprisingly) entitled "Mathematical Biophysics" (Rashevsky, 1960), a book that was revised and reedited more than once (most of the above mentioned studies can be found in this publication). The original edition, dating back to 1938, covered cellular biophysics, excitation phenomena and the central nervous system, with emphasis in the physical representation (Morowitz, 1965). Subsequently, Rashevsky delved into still more abstract mathematical approaches, as I describe later.

I wish to point out that the initial efforts of Rashevsky are both important and largely unrecognized by contemporary philosophers and historians of science. Indeed, the academic infrastructure and the research agenda established by this forerunner were crucial in subsidizing the development of important scientific contributions made by other researchers. Perhaps the most unexpected case is the crucial role played by Rashevsky in the line of investigation leading to the celebrated McCulloch-Pitts model of neural networks, published in 1943. Consider the following statement, which nicely summarizes a commonly held belief of the scientific and philosophical communities: "The neural nets branch of AI began with a very early paper by Warren McCulloch and Walter Pitts..." (Franklin, forthcoming). Actually, the first mathematical descriptions of the behavior of "nerves" and networks of nerves are to be credited to Rashevsky, who during the early 1930´s published several papers concerning a mathematical theory of conduction in nerves, based on electrochemical gradients and the diffusion of substances (Abraham, 2002). Rashevsky´s fundamental idea was to use two linear differential equations together with a nonlinear threshold operator (Rashevsky, 1933). It was in this paper, hence, that Rashevsky´s "two-factor" theory of nerve excitation became public for the first time. This theory was based on the diffusion kinetics of excitors and inhibitors (Abraham, 2002). Only many years later the unknown "substances" were correctly identified as concentrations of sodium and potassium, thanks to the important work of Hodgkin and Huxley (Cull, 2007). I am not willing to get into the technical details here (a thorough investigation will be published elsewhere), but suffice it to say that Rashevsky argued that his simple mathematical model could fit empirical data, available at the time, regarding the behavior of single neurons. Still more important, he postulated that these model neurons could be connected in networks, in order to yield complex behavior, and even allow the modeling of the entire human brain (Cull, 2007). Latter, close to the end of that decade, Walter Pitts was introduced to Rashevsky by Rudolf Carnap, and accepted into his mathematical biology group (Cowan, 1998). Together, Pitts (who was a superb mathematician) and the "philosophical psychiatrist" Warren S. McCulloch (Abraham, 2002) published their groundbreaking paper, entitled "A Logical Calculus of the Ideas Immanent in Nervous Activity" (McCulloch and Pitts, 1943). Many years later, McCulloch recalled that he and Pitts were able to publish their ideas in Rashevsky´s journal thanks to Rashevsky´s defense of mathematical and logical ideas in the field of biology (Abraham, 2002). The objective fact that Rashevsky was apparently the first investigator to come up with the idea of a "neural net" mathematical model (Rosen, 1991), however, was largely neglected.

According to Rosen (1991), by the 1950s Rashevsky had explored many areas of theoretical biology, but he felt that his approach still lacked "genuinely fresh insights." Thus, he suddenly took a wholly new research direction, turning from mathematical methods closely associated with empirical data to an overarching search for general biological principles. Putting aside Rosen´s opinion, this radical turn is seemingly most easily justifiable simply as a strong reaction to Rashevsky´s own critics, who claimed that his mathematical biophysics approach was not a novelty anymore (ironically) –, given that many researchers had already began incorporating models derived from physics, as well as quantitative methods, in their work (Cull, 2007).

Anyway, the fact is that Rashevsky expressed a bunch of novel ideas in a paper interestingly entitled "Topology and Life: In Search of General Mathematical Principles in Biology and Sociology". In the paper, after pointing out most major developments in the mathematical biology of the time, he goes on saying: "All [these] theories ... deal with separate biological phenomena. There is no record of a successful mathematical theory which would treat the integrated activities of the organism as a whole." (Rashevsky, 1954, p. 319-320). According to him, it was important to have the knowledge that diffusion drag forces are responsible for cell division and that pressure waves are reflected in blood vessels, as well as to have a mathematical formalism for dealing with complicated neural networks. But then he emphasized that there was nothing so far in these theories indicating that an adequate functioning of the circulatory system was fundamental for the normal operation of intracellular processes; Furthermore, there was nothing in the formalisms showing that an elaborated process in the brain, that resulted, e.g., in the location of food, was causally connected with metabolic processes going on in the cells of the digestive system. The same was true regarding the causal nexus between a failure in the normal behavior of a network of neurons and the cell divisions that resulted from a stimulation of the process of healing due to the accidental cutting of, say, a thumb (Rosen, 1991). And yet, according to him, "this integrated activity of the organism is probably the most essential manifestation of life." (Rashevsky, 1954, p. 320). Unfortunately, Rashevsky argued, one usually approaches the effects of these diffusion drag forces simply as a diffusion problem in a *specialized physical system*, and one deals with the processes of circulation simply as special hydrodynamic problems. Hence, the "fundamental manifestations of life" are definitely excluded from all those biomathematical theories. In other words, biomathematics, according to Rashevsky, lacked a capacity to adequately describe true *integration* of the parts of any organic system. As a result, it was useless to try to apply the physical principles, used in the aforesaid mechanical models of biological phenomena, to develop a comprehensive theory of life. Similar lines of criticism could be applied, I submit, to modern theoretical frameworks attempting to provide integrated models of the living organism (or the brain itself). This line of thought was further developed by Rashevsky´s student Robert Rosen, who yielded an interesting theoretical framework, built around a special notion of complexity. He also promoted the use of new mathematical tools, like category theory, in theoretical biology investigations (Rosen, 1991, 2000).

  According to Rashevsky, putting aside the possibility of constructing a physicomathematical theory of the organism *based on the physicochemical dynamics of cells and of cellular aggregates* does not prevent one from trying to find alternative pathways. What are the possibilities, then? The key to understanding Rashevsky's perspective, I suggest, is to start analyzing some of his fundamental premises. In fact, Rashevsky used to believe that the biomathematics of his time was in a *pre-Newtonian stage of development*, despite the elaborate theories then available. In pre-Newtonian physics there existed simple mathematical treatments of isolated phenomena, but it was only with the arrival of Newton´s principles, incorporated in his laws of motion, that physics attained a more comprehensive and unified synthesis. Ordinary models of biomathematics, like the models of theoretical physics, are all based in *physical principles*. But Rashevsky seemed to suggest that they should instead be based on genuinely *biological principles*, in order to capture the *integrated activities of the organism as a whole* (Rosen, 1991). This is apparent in his words: "We must look for *a principle which connects the different physical phenomena* involved and expresses the biological unity of the organism and of the organic world as a whole." (Rashevsky, 1954, p. 321, italics added). He also argued that mathematical models are transient in nature, while a general principle, when discovered, is perennial. For example, it is possible to devise several alternative models, all obeying the laws of Newton (one model being e.g. the "billiard ball" molecule of the kinetic theory of gases). On the same guise, there are distinct cosmological models, all based upon Einstein's fundamental principles (Peacock, 1999; Dalarsson and Dalarsson, 2005). It is clear then that Rashevsky wished to conceive general principles, in biology, enjoying the same status earned by principles in theoretical physics. After all, he was trained as a theoretical physicist.

## 3. Looking for general biological principles

I already highlighted the fact that Rashevsky was a researcher that eagerly pursued general biological principles, and he indeed explicitly managed to propose some. In what follows, I briefly examine (following Rosen, 1991) the nature of some of these principles. The first principle I would like to emphasize is Rashevsky's *principle of adequate design of organisms*, originally denominated *principle of maximum simplicity*, and introduced as early as in 1943. As originally formulated it states that, given that the same biological functions can be performed by different structures, the particular structure found in nature is the *simplest* one compatible with the performance of a function or set of functions. The principle of maximum simplicity applies therefore to different *models* of mechanisms, of which the simplest one is to be preferred. But given that simplicity is a vague notion in this case, being difficult to find out a measurement standard, Rashevsky self-critically turned to a slightly different version, denominated *principle of optimal design*. In this case, it is required that a structure necessary for performing a given function be *optimal* relatively to energy and material needs. But it can be argued that there is still some imprecision here, because a structure that is optimal with respect to material needs is not necessarily optimal as far as energy expenditures are concerned. Hence, a more straightforward notion was in need. Accordingly, Rashevsky turned to the last formulation of his principle, this time putting aside the notion of optimality:

> When a set of functions of an organism or of a single organ is prescribed, then, in order to find the shape and structure of the organ, the mathematical biologist must proceed just as an engineer proceeds in designing a structure or a machine for the performance of a given function. The design must be *adequate* to the performance of the prescribed function under specified varying environmental conditions. This may be called the *principle of adequate design of the organism*. (Rashevsky, 1965, p. 41, italics added).

I note that the notion of "adequate" is still also a bit vague, but I am not pursuing this discussion further in this paper. Let me instead raise a more interesting question, which repeatedly arises in philosophy of biology, particularly in those areas of inquiry closely associated to the Darwinian theory of biological evolution: Is Rashevsky´s principle of adequate design of the organism teleological? To this objection Rashevsky himself had an answer: all variational principles in physics are "teleological" or "goal-directed," beginning with the principle of least action (see Lanczos, 1970, for a detailed mathematical account of this and other physical principles). Other investigators subsequently offered similar justifications, e.g. Robert Rosen, who dedicated a full chapter of his book "Essays on Life Itself" (Rosen, 2000), entitled "Optimality in Biology and Medicine", to this technical discussion. Another objection that Rashevsky was well aware of was that the principle of adequate "design" seemed to imply some sort of creative intelligence. Must one, in this case, presuppose a "universal engineer" of sorts? Not necessarily, because, like most scientific principles, the *principle of adequate design of the organism* "offers us merely an operational prescription for the determination of organic form by calculation." (Rashevsky, 1965, p. 45). Here, I suggest that one would perhaps do best simply not employing the term "design," that also seems to be rather misguiding in this context. On the other hand, according to Rashevsky, the principle could perhaps follow directly from the operation of natural selection, which would only preserve "adequate" organisms, although it could turn out to be an independent principle. This too, I suggest, could be nourishment for heated discussions among contemporary philosophers of biology.

Perhaps still more thought-provoking is Rashevsky´s "principle of biological epimorphism," that emphasizes qualitative relations as opposed to quantitative aspects, topology instead of metrics. It can be argued that a given biological property in a higher organism has many more elementary processes than the equivalent biological property of a lower one. Examples of biological properties are perception, locomotion, metabolism, etc. The principle is based upon the fact that different organisms can be epimorphically mapped onto each other, after the biological properties were already clearly distinguished and represented. In such epimorphic mappings, the basic *relations* characterizing the organism as a whole are preserved. Given Rashevsky's mathematical proclivities, he wanted to put his principle into a precise and rational context. Among the several branches of relational mathematics, topology reigns supreme. Before going on, I think it necessary to briefly digress about this topic.

It is a known fact that topological ideas are present in most branches of modern mathematics. In a nutshell, topology is the mathematical study of properties of objects, which are preserved through deformation, stretching and twisting (tearing is forbidden). Hence, one is entitled to say that a circle is topologically equivalent to an ellipse, given that one can be transformed into the other by stretching. The same is valid for a sphere, which can be transformed into an ellipsoid, and vice versa. Topology has indeed to do with the study of objects like curves, surfaces, the space-time of Minkowsky (in relativity theory, see Peacock, 1999), physical phase spaces and so on. Furthermore, the objects of topology can be formally defined as "topological spaces." Two such objects are homeomorphic if they have the same topological properties. Using such perspective, Rashevsky postulated that to each organism there is a corresponding topological "complex." More complicated complexes correspond to higher organisms, and different complexes are converted into each other by means of a universal rule of geometrical transformation. Furthermore, they can be mapped onto each other in a many-to-one manner, preserving certain basic relations. Rashevsky expressed his "principle of biological epimorphism" by postulating that, if one represents geometrically the relations between several functions of an organism in a single convenient topological complex, then the topological complexes that represent different organisms are obtainable, via a proper transformation, from just one or a few primordial topological complexes. A previous version of this principle is what Rashevsky used to call the "principle of bio-topological mapping." According to this principle, the topological complexes by means of which diverse organisms are represented are all obtainable from one or a few primordial complexes *by the same transformation*. This transformation contains one or more parameters, different values of it corresponding to different organisms (Rashevsky, 1954). The considerations above may hopefully give us a glimpse of Rashevsky's relational approach to the study of life, epitomized in the expression "relational biology," that he coined in order to help delineate a clear framework for thinking in the life sciences.

## 4. Conclusion

Nicolas Rashevsky (who passed away in 1972) was a pioneer in theoretical biology, having inaugurated the school of "mathematical biophysics" and subsequently pioneered the field of "relational biology" or (still another term that he coined) "biotopology." I call attention to the fact that the latter must definitely be distinguished from "topobiology," a term coined by Nobel laureate Gerald Edelman in the context of cell and embryonic development research. Edelman´s theory postulates that differential adhesive interactions among heterogeneous cell populations drive morphogenesis, and explains, among other things, how a complex multi-cellular organism can arise from a single cell (Edelman, 1988).

I already emphasized that the creation of "The Bulletin of Mathematical Biophysics" was an important tool for establishing and helping broadcast Rashevsky´s work (as well as the proposals of his own students). Furthermore, I pointed out that the work of Rashevsky implies that at least some aspects of contemporary theoretical biology and neuroscience have older roots than previously thought. This is exemplified by Rashevsky´s active role in the body of research that paved the way to the development and publication of the McCulloch-Pitts model of neural networks. Finally, I suggested that Rashevsky´s criticism of purely mechanical and non-integrative approaches to biology may as well be evaluated under the light of current theories, including proposals in theoretical biology – and, once again, in neuroscience. The critical analysis of these claims, I submit, is an interesting and yet largely unexplored subject-matter to philosophers of science.

Rashevsky´s influence still reverberates in important scientific research areas such as neural networks and non-equilibrium pattern formation, among others. However, as I see it, relational biology effectively came of age with the far more encompassing and methodical work of Rashevsky's former student Robert Rosen (who passed away in 1998 – see Rosen 1991, 2000) and of his followers. A very active contemporary player worth mentioning is a bright pupil of Rosen, the mathematical biologist Aloisius H. Louie (see Louie, 2009, 2013).

**Acknowledgments**

Some ideas presented in this paper first came to my mind during the development of my Ph.D. dissertation, years ago, and ended up as part of a chapter (still unpublished). Accordingly, I would like to thank CNPq for the support at the time, by means of a much needed research grant.

**References**

ABRAHAM, T. H. (Physio)logical circuits: The intellectual origins of the McCulloch-Pitts neural networks. *Journal of the History of the Behavioral Sciences*, v. 38, p. 3-25, 2002.

COWAN, J. D. Interview with J. A. Anderson and E. Rosenfeld. In: Anderson, J. A.; Rosenfeld, E. (Eds.), *Talking nets: An oral history of neural networks*. Cambridge MA: MIT Press, 1998. Pp. 97-124.

CULL, P. The mathematical biophysics of Nicolas Rashevsky. *BioSystems*, v. 88, p. 178-184, 2007.

DALARSSON, M., DALARSSON, N. *Tensors, relativity and cosmology*. New York: Elsevier, 2005.

EDELMAN, G. M. *Topobiology: An introduction to molecular embriology*. New York: Basic Books, 1988.

FISHER, R. A. *The genetical theory of natural selection.* Oxford: Clarendon, 1930.

FISHER, R. A. *The theory of inbreeding.* London: Oliver and Boyd, 1949.

HALDANE, J. B. S. *A mathematical theory of natural and artificial selection.* Cambridge: Cambridge University Press, 1924.

HINSHELWOOD, C. N. *The chemical kinetics of the bacterial cell*. Oxford: Clarendon, 1946.

KOSTITZIN, V. A. *Biologie mathématique.* Paris: Librairie Armand Colin, 1937.

LANDAHL, H. D. A kinetic theory of diffusion forces in metabolizing systems. *Bulletin of Mathematical Biophysics*, v. 4, p. 15-26, 1942a.

LANDAHL, H. D. A mathematical analysis of elongation and constriction in cell division. *Bulletin of Mathematical Biophysics*, v. 4, p 45-62, 1942b.

LANCZOS, C. *The variational principles of mechanics.* Toronto: University of Toronto Press, 1970.

LOTKA, A. J. *Elements of physical biology.* Baltimore: Williams and Wilkins, 1925.

LOUIE, A. H. *More than life itself: A synthetic continuation in relational biology*. Frankfurt: Ontos-Verlag, 2009.

LOUIE, A. H. *The reflection of life: Functional entailment and imminence in relational biology.* New York: Springer, 2013.

McCULLOCH, W. S., PITTS, W. A logical calculus of the ideas immanent in nervous activity. *Bulletin of Mathematical Biophysics*, v. 5, p. 115-133, 1943.

MOROWITZ, H. J. The historical background. In: Waterman, T. H.; Morowitz, H. J. (Eds.), *Theoretical and mathematical biology*. New York: Blaisdell, 1965. pp. 24-35.

ODUM, H. T. *Systems ecology: An introduction.* New York: John Wiley & Sons, 1983.

PEACOCK, J. A. *Cosmological physics*. Cambridge: Cambridge University Press, 1999.

RASHEVSKY, N. Outline of a physico-mathematical theory of excitation and inhibition. *Protoplasma*, v.20, 1933.

RASHEVSKY, N. The mechanism of cell division. *Bulletin of Mathematical Biophysics*, v.1, p. 23-30, 1939.

RASHEVSKY, N. Topology and life: In search of general mathematical principles in biology and sociology. *Bulletin of Mathematical Biophysics*, v.16, p. 317-348, 1954.

RASHEVSKY, N. *Mathematical Biophysics*. Chicago: University of Chicago Press, 1960.

RASHEVSKY, N. Models and mathematical principles in biology. In: Waterman, T. H.; Morowitz, H. J. *Theoretical and mathematical biology*. New York: Blaisdell, 1965. p. 36-53.

ROSEN, R. *Life itself: A comprehensive inquiry into the nature, origin and foundation of life.* New York: Columbia University Press, 1991.

ROSEN, R. *Essays on life itself*. New York: Columbia University Press, 2000.

SHANNON, C. E. A mathematical theory of communication. *Bell System Technical Journal*, v.27, p. 379-423, 1948.

SCHRÖDINGER, E. *What is life?* Cambridge: Cambridge University Press, 1944.

SMOCOVITIS, V. B. *Unifying biology: The evolutionary synthesis and evolutionary biology.* Princeton: Princeton University Press, 1996.

FRANKLIN, S. History, motivations, and core themes. In: Frankish, K.; Ramsey, W. M. (Eds.), *The Cambridge handbook of artificial intelligence*. Cambridge: Cambridge University Press, forthcoming.

THOMPSON, D'A. W. *On growth and form.* Cambridge: Cambridge University Press, 1917.

ULANOWICZ, R. E. An hypothesis on the development of natural communities. *Journal of Theoretical Biology*, v. 85, p. 223-245, 1980.

ULANOWICZ, R. E. *Ecology, the ascendent perspective.* New York: Columbia University Press, 1997.

VOLTERRA, V. *Leçons sur la théorie mathématique de la lutte pour la vie.* Paris: Gauthier-Villars, 1931.

WIENER, N. *Cybernetics.* New York: MIT Press, 1948.